# Reputation-based partition scheme for IoT security

Zhikui Chen[1] | Muhammad Zeeshan Haider[1] | Naiwen Luo[1] | Shuo Yu[2] | Xu Yuan[1] | Yaochen Zhang[3] | Tayyaba Noreen[1]

[1]School of Software, Dalian University of Technology, Liaoning, Dalian, China

[2]School of Computer Science and Technology, Dalian University of Technology, Liaoning, Dalian, China

[3]Inspur Software Technology Co. LTD, Shandong, Jinan, China

**Correspondence**
Shuo Yu, School of Computer Science and Technology, Dalian University of Technology, Dalian 116024, China.
Email: shuo.yu@ieee.org

Xu Yuan, School of Software, Dalian University of Technology, Dalian 116620, China.
Email: david@dlut.edu.cn

**Funding information**
China's National Key Research and Development Program, Grant/Award Number: 2018YFC0831305

**Abstract**

With the popularity of smart terminals, such as the Internet of Things, crowd-sensing is an emerging data aggregation paradigm, which plays a pivotal role in data-driven applications. There are some key issues in the development of crowdsensing such as platform security and privacy protection. As the crowd-sensing is usually managed by a centralized platform, centralized management will bring various security vulnerabilities and scalability issues. To solve these issues, an effective reputation-based partition scheme (RSPC) is proposed in this article. The partition scheme calculates the optimal partition size by combining the node reputation value and divides the node into several disjoint partitions according to the node reputation value. By selecting the appropriate partition size, RSPC provides a mechanism to ensure that each partition is valid, as long as the maximum permissible threshold for the failed node is observed. At the same time, the RSPC reorganizes the network periodically to avoid partition attacks. In addition, for cross-partition transactions, this paper innovatively proposes a four-stage confirmation protocol to ensure the efficient and safe completion of cross-partition transactions. Finally, experiments show that RSPC improves scalability, low latency, and high throughput for crowdsensing.

## 1 | INTRODUCTION

The rapid development of information and communication technology has facilitated the development of traditional computer-aided industries toward intelligent industries characterized by data-centric decision-making.[1] The Internet of Things (IOT) has played a key role in this process. The IOT interfaces the physical industrial environment and the cyberspace of the computing system to form a cyber-physical framework. It seeks to enhance operating efficiency and capacity, as well as minimize machine downtime and improve quality of product.[2] Manufacturing, logistics, the food sector, and public utilities are among the industries that it can benefit. However, the existing IoT system still uses a centralized architecture. The security problem of the centralized architecture is that all data information security depends on the central server, which can easily cause the collapse of the entire centralized architecture.[3] For example, hackers maliciously attack the core node server in the IoT system, steal user data and business secrets, disrupt the normal operation of the core server, spread and illegally operate business secrets and user personal information, and disrupt the service support of the IoT, which will affect the public Property security and personal information security are extremely harmful.

Therefore, in terms of information security and data storage interaction, the traditional IoT has been unable to meet the increasing access of a massive number of IoT terminals. Blockchain technology provides a decentralized alternative for large-scale collaboration between IoT devices because it is non-tamperable, secure, and traceable,[4,5,6] which solves the urgent need for security in the growth of the IoT, and fundamental components of Blockchain technology are decentralized trust enabled by a consensus process and distributed storage enabled by a tamper-proof ledger. These capabilities can help any application with various stakeholders since they allow for transparent interactions without the essential for a trusted third party. These distinguishing characteristics of BC make it attractive for enabling distributed security and privacy on the IoT. The decentralization in IoT terminal equipment undoubtedly provides the best space for decentralized architecture.[7]

Blockchain is essentially a distributed shared ledger technology where multiple parties participate in the joint maintenance. It is built on cryptography, distributed systems, network protocols and other technologies to achieve decentralized trusted value transmission.[8,9] The development of blockchain technology in the recent years has opened new possibilities in a variety of fields, such as IoT, supply chain, copyright protection, social networking, and medical treatment.[10,11,12]

However, the existing blockchain-based systems face some challenges in terms of scalability. For example, all transactions must be processed by each node on the block-chain network, and all data of the whole network, and the processing capacity of the entire blockchain system is limited by the processing capacity of a single computing node. Secondly, as the number of nodes grows, the consensus algorithm is affected, the overall processing capacity of the system not only does not increase, but also even decreases.[13] Existing blockchain systems, such as Bitcoin uses Proof of Work (PoW) consensus, which processes seven transactions per second, on the other hand Ethereum uses the continuously improved proof of stake (PoS) consensus,[14] which processes 20 transactions per second. The hyperledger block-chain uses the practical byzantine fault tolerance (PBFT)[15] developed by IBM to reach a consensus and handles 3500 transactions per second. Obviously, compared with centralized databases, the scalability of blockchain technology has become the biggest bottleneck for blockchain applications. In addition, typical Bitcoin and Hyperledger Fabric require at least 1 GB and 4 GB of memory, respectively, but most IoT devices have KB–MB of memory.[16] A promising way to break this dilemma is to use partitioning technology. Partitioning has the function of parallelizing transaction processing, which can make the blockchain more scalable and improve throughput.[17]

Through the partition technology, a variety of methods to extend the blockchain have been proposed. It is worth noting: ELASTICO,[18] OmniLedger,[19] RapidChain,[20] Monoxide.[21] However, the above research did not answer the following questions well:

1. How to detect and evaluate the reliability of IoT nodes to ensure that credible committee members are selected to ensure the stability of the consensus network?
2. How to figure out the best partition size for the IoT blockchain to have the best storage performance of the entire network increases linearly with the increase of IoT nodes?
3. How can the security and efficiency of cross-partition transactions be improved?

In response to the above-discussed problems, this article proposes a reputation-based partition scheme (RSPC) suitable for physical networks. Firstly, RSPC evaluates the reputation of nodes based on the cutting-edge work of this article (Yuan et al. 2021). Secondly, the partition scheme combines the reputation value to calculate the consensus security size and consensus complexity to determine the optimal partition size. Then, RSPC adopts a multi-chain structure, and each node only stores the partition chain in its own partition and the global chain that only contains UTXO state backup. The goal of synchronously expanding the storage performance of the blockchain with the increase of node size is achieved. In addition, to reduce data migration overhead while ensuring system security, RSPC regularly reorganizes each partition. Before reorganizing the network, each district leader node updates the UTXO status on the global chain. After reorganization, each node starts a new partition cycle based on the latest global chain, thereby avoiding the data migration overhead that occurs when reorganizing the network. Finally, for cross-partition transactions, this paper innovatively proposes a 4-stage confirmation protocol to improve the security and efficiency of processing cross-partition transactions. In other words, RSPC increases the throughput of the consensus network and increases the scalability of the blockchain while guaranteeing that the whole consensus process satisfies the requirements of the validity, consistency, and integrity of the consensus protocol. Protection can be ensured as the network grows more powerful.

The data of the IoT devices cannot be tampered with, and the network is safe, reliable, and efficient. The following are the article's primary innovations:

1. A partition algorithm combining reputation score and partition-based blockchain is designed. The algorithm is based on the node reputation value through the Kademlia and PRNG random number generator algorithm to assign nodes to different consensus partitions that ensures the efficiency and randomness of the partition.
2. The multi-chain structure of partition chain and global chain is designed. The goal of synchronously expanding the storage performance of the blockchain with the increase of node size is achieved.
3. A new cross-partition trading algorithm is designed. The algorithm is based on the global leader elected by each district to confirm the involved districts in both directions, so that cross-division transactions can be carried out efficiently and safely.

The rest of this article is organized in the following manner. In the second section, we reviewed the technical status of the partitioning algorithm. Third section provides the background knowledge of the EBRC agreement. In the fourth section, the system model is introduced. Fifth section describes the implementation and performance evaluation of the proposed partitioning algorithm is carried out, and the attack types are discussed, and analyzed the security of our protocol. Finally, we give the conclusion of this article in Section 6.

## 2 | RELATED WORK

Crowdsensing gives users access to the sort of cloud platform, social traits, and other data in addition to allowing users to participate on massive awareness activities. Similar to interpersonal connections, the degree of trust influences how we manage privacy in the IoT ecosystem with blockchain decentralization. Increasingly data will be shared between people, businesses, governments, and ecosystems as IoT nodes become more linked. The reliability of sensors, devices, computers, and cloud connections depends greatly on those linkages. The threat to established systems will increase with the connection of more IoT device types (similar to adding intrusion points), raising overall security risks. It is challenging to build confidence with an IoT system or device if privacy settings cannot be restricted.[22,23] Here comes the blockchain to rescue the IOT crowdsensing framework as there is a lot recent research on the blockchain-based IOT crowdsensing platforms like Elastico[18] is the first partition-based public chain consensus protocol, and proposed a PoW-based network partition method to extend the IOT transaction verification of public blockchains. Through partitioned networks and transactions in an IOT crowdsensing environment, it is able to verify transactions in parallel. PoW is used by ELASTICO, but this method does not address the problem of energy consumption or other PoW-related problems. A cross-partition transaction solution is not provided by Elastico, which also randomly chooses nodes when splitting the network and merely provides probabilistic accuracy without a way to differentiate malicious partitions.[13] In contrast, in the presence of malicious and faulty nodes, our technique presented a methodology for determining the appropriate fragment size based on consensus security and complexity.[24]

OmniLedger[19] proposed a partitioning protocol to solve some limitations of ELASTICO. It uses a probability model RandHound, which is a distributed randomness generating mechanism,[25] and the VRF leader election algorithm[26,27] to partition the network. The partition evaluates the request through a variant of the PBFT consensus protocol and responds to the client with an accepted or rejected status. However, OmniLedger has a large amount of data migration consumption in its regular partition reorganization strategy. In addition, the partition is likely to be malicious (ie, composed of most malicious nodes) in IOT environment. Although it is small, it does not propose any technology to detect malicious partitions.[13]

RapidChain[20] does not use any public key infrastructure, instead requiring each node to solve the PoW issue before proceeding to the next round of consensus. Before consensus, the network is divided into multiple smaller node groups (ie, network fragments), and each node manages its own ledger and processes a set of disjointed transactions. Therefore, it can parallelize transaction processing and storage allocation by partitioning the network and workload IOT architecture. However, the RapidChain partitioning scheme only emphasizes the security aspect and does not consider the overall system performance.

Monoxide[21] is a blockchain partition system based on PoW. It recommends partitioning the network into several asynchronous consensus zones, each of which would function as its own PoW blockchain. Monoxide demonstrated higher throughput in the blockchain network but did not explain the partitioning scheme that provides the best performance. Furthermore, miners still overcome PoW difficulties in order to submit new blocks, reducing their energy efficiency in crowdsensing environment.

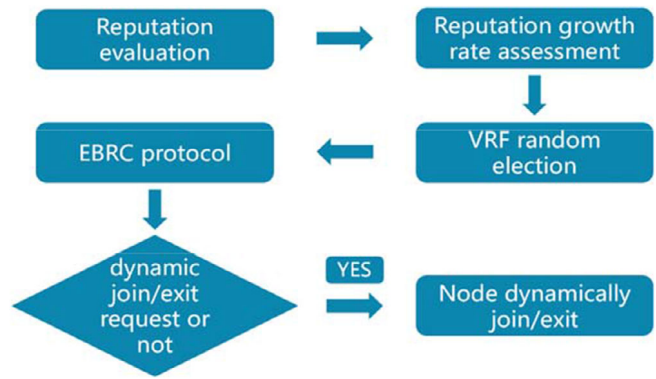

**FIGURE 1** Basic consensus process of EBRC

SSChain[28] divides the blockchain network into a root chain network and a partitioned network. The hash power in the root chain network and the blockchain network will be dynamically adjusted by the market incentive mechanism, so that the hash power is evenly distributed in different partitions. Nodes can freely choose to join the partition with higher values. A node can belong to the root chain network and the partition network at the same time. The root chain network is responsible for verifying the partition blocks over a period to generate the root block, while saving the complete partition area Block and root block. However, similar to ELASTICO, SSChain's two consensuses both use POW, and its transaction processing performance is low, and the consensus process can only guarantee final consistency rather than strong consistency, which is easy to cause the blockchain to fork, Transaction confirmation delays and other issues. Zhang[29] et al. proposed a hierarchical structure based on "groups." Divide all nodes into several groups, and each group has a master node. Hierarchical consensus is divided into two stages: local and global. First, the client node broadcasts the request to all nodes, and each group has a local consensus of the request within the group; then, the master node of each group performs a global consensus again, and the entire request is considered complete.

Feng et al.[30] proposed a SDMA-PBFA, a scalable dynamic multi-agent hierarchical PBFT technique that can shorten communication messages in the consensus process from $O\left(n^2\right)$ to $O(nklogkn)$. The basic concept is to split the nodes into several groups, and each group selects a node as an agent for IOT devices. The client node request is first distributed to all agents, and the agents complete the consensus process within the group in a permissioned blockchain network. Finally, each agent informs the client node that the request is completed.

Usually, network partitioning is considered as an on-chain performance extension technology. When a network has a particular number of nodes, the more network partitions there are, the fewer nodes on the chain there are inside each partition. This can increase the efficiency of single-chip consensus and the overall network's transaction efficiency; however, the blockchain network's security will be compromised. On the contrary, when partitioning, consider such a scenario. That is, when the system allocates too many malicious nodes to the same partition, causing the partition's proportion of malicious nodes to exceed the safety threshold (such as 1/3 of the Byzantine consensus), and the partition cannot reach a consensus and the partition becomes invalid. As the number of network nodes grows, determining the optimum partitioning technique to ensure that each partition consensus does not fail becomes an issue that should be quickly resolved.

Therefore, this work provides a mathematical method for determining the ideal partition size based on reputation value while also considering performance and node reliability in IOT crowdsensing architecture. In addition, the partition chain and global chain multi-chain structure that enables the protocol to be expanded in proportion to the number of nodes, methods to reduce the data migration overhead when reorganizing the partition network, and algorithms for efficient and safe cross-partition transactions are also designed.

## 3 | PRELIMINARIES

Our work is a network partitioning scheme based on the EBRC (Efficient Byzantine Consensus Mechanism Based on Reputation) consensus algorithm.[31] Therefore, we give a brief explanation of the mechanism in this section. Figure 1 shows the fundamental consensus mechanism.

The EBRC consensus algorithm is improved algorithm based on the classic PBFT. The algorithm first comprehensively evaluates the node's reputation value and the growth rate of the reputation value, determines the election authority of the node in the system, and ensures the reliability of the node. Secondly, a node election algorithm based on verifiable random function (VRF) is designed, which randomly selects nodes from the nodes with voting rights to become consensus nodes. Then, establish a penalty mechanism for malicious nodes. When a node produces malicious behavior, the reputation of the node will decline. EBRC will correspondingly impose large margin penalties on the system, thereby reducing the probability of malicious nodes attacking the IoT system again.[31] In addition, DJEP, dynamic joining and exiting protocol that can significantly enhance the dynamic change characteristics of nodes in the blockchain system, is designed. Finally, it includes an efficient reputation-based Byzantine negotiation mechanism. The Byzantine negotiation mechanism is divided into two stages: preparation and submission. In the preparation phase, each slave node confirms the consensus request initiated by the master node. If the master node is found to be evil, it can initiate a view conversion request, update the reputation value according to the reputation model, deduct the deposit of the master node and replace the master node. If a correct slave node gets $2f + 1$ valid PREPARE messages at the end of the first phase, then it proceeds to the commit phase. The node broadcast the submission message to other nodes. When node receives a message such as $2f + 1$, it knows that it is ready to submit. The node then executes the operation to collect the result and responds to the client with a message. If the client receives $f + 1$ matching response messages, then the re-quest is considered processed, and the returned result is accepted. Finally, update the current node reputation value and enter the next round of consensus.

Similar to the classic PBFT consensus mechanism, EBRC satisfies the fault tolerance condition of the Byzantine consensus, that is,

$$f \geq \frac{n-1}{3} \quad (1)$$

## 4 | RSPC PROTOCOL

### 4.1 | Protocol overview

In response to the issues raised in Section 2, this work is based on EBRC and proposes a RSPC suitable for the IoT, as shown in Figure 2. The protocol is split into three parts: optimal partition size calculation, node allocation, and consensus.

### 4.2 | Optimal partition size calculation

Optimal partition size calculation: At the beginning of each epoch, In the presence of malfunctioned or faulty nodes, the optimal partition size is calculated based on the theoretical consensus security size and consensus difficulty. By selecting the correct partition size with the best overall performance, the effectiveness of the partition is guaranteed. Among them, the malicious or faulty node defined in this work is the reputation value $R \leq 0.2$, $R \in [0, 1]$ (the initial value is 0.5), which is calculated by the reputation module in the EBRC consensus. Please refer to Section 4 and three subsections.

### 4.3 | Node allocation

At the beginning of each epoch, all network nodes are randomly allocated to different partitions based on the node reputation value to ensure the efficiency and randomness of the partitions. Please refer to Section 4.3.

### 4.4 | Consensus

Consensus is divided into intra-regional consensus and cross-regional consensus.

1. *Intra-partition consensus*: In order to prevent transactions from being reprocessed in different chunks, the partition to which the transaction belongs is assigned based on the sender address. The client sends a transaction (transaction)

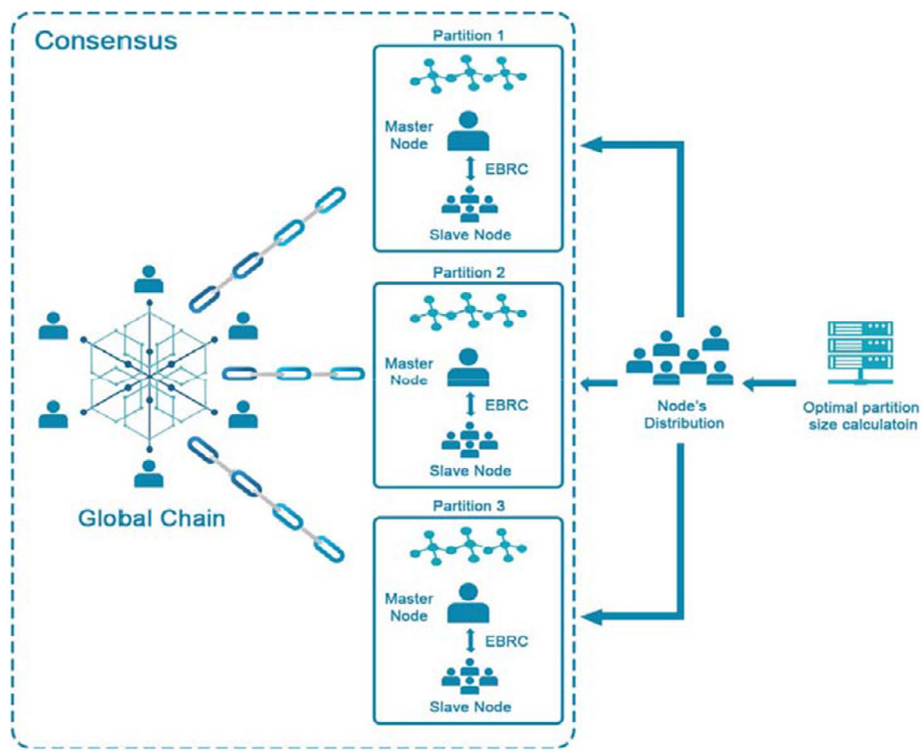

**FIGURE 2** RSPC protocol architecture

TX to the partition responsible for entering UTXOS (unspent transaction output). All partitions run EBRC consensus in parallel to process sub-blocks generated by network transactions, which are broadcast by the main node on the chain within each partition, and each partition owns the partition chain. Thus, the transaction processing and storage performance of blockchain are expanded synchronously with the size of nodes. And according to the behavior of all verifiers in a consensual way, calculate and achieve a consensus on the credibility score. Reputation scores enhance the reliability of consensus by helping to calculate the best partitions and to select highly competent leaders within the partitions. After that, the node synchronizes and updates the stored credit scores based on these credit values from all partitions, and the system starts a new round (Section 4.3).

2. *Cross-partition consensus*: If the sender address and receiver address are in the same partition, then they are directly packaged into the block. If the sender and receiver are not in the same partition, then cross-region transactions will occur. In order to maximize the degree of concurrency of transactions, this paper designs a global leader node elected by each district to complete the consensus through a four-phase two-way confirmation protocol. Please refer to Section 4.3.

Therefore, according to the RSPC composed of the above three parts, while ensuring that the overall consensus process meets the requirements of the validity, consistency and integrity of the consensus protocol, it also increases the throughput of the consensus network and enhances the scalability of the blockchain. In addition, in order to eliminate the overhead of data migration while maintaining system security, RSPC adopts a multi-chain structure of partition chain and global chain, so that the blockchain network storage performance increases synchronously with the increase of nodes and avoids the occurrence of partition reorganization. Data migration overhead.

## 4.5 | Optimal partition size calculation

The blockchain network nodes are partitioned, and the pending transactions are distributed to each partition, so that each partition verifies a set of unrelated transactions, thereby ensuring the scalability of the blockchain. Although the partition system results in higher scalability, it also demands the deployment of additional security and fault tolerance mechanisms in the case of malicious or malfunctioning nodes in the network. As we discussed in Section 2, many methods have been proposed to partition blockchains with different security and fault tolerance models. However, there is no literature that comprehensively considers the partitioning scheme from the three aspects of performance, safety and fault tolerance. The calculation of the optimal partition size is the key to RSPC, which dynamically adjusts the committee size of the entire network. Therefore, this work determines the optimal partition size based on the consensus security size and consensus complexity when there are malicious and faulty nodes, which is also called the committee size.

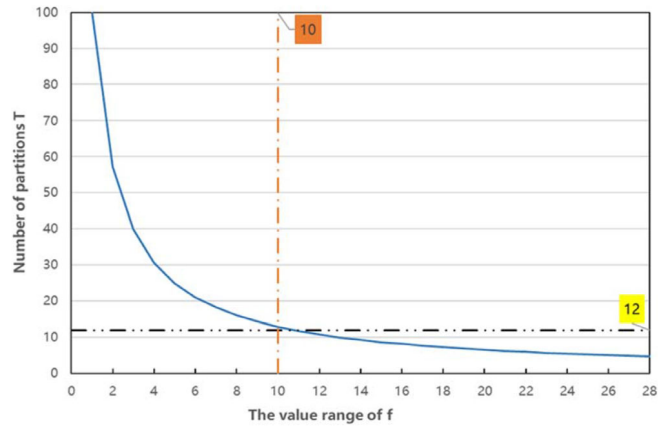

**FIGURE 3**   Consensus security size (when $N = 400$, for different $f$, the value range of $T$)

First, consider the size of consensus security. Kantesariya et al.[32] determined the partition size based on the PBFT protocol and its variants, but the partition size calculated in the algorithm does not meet the conditions for reaching a consensus. For example, according to the algorithm, the division size is 60 when the number of nodes is 100. With 1 node in each partition, it is impossible to reach a PBFT consensus. So, this work improves it as follows. The N-node blockchain network (except the network leader) is divided into T disjoint partitions called committees, and the size of each partition $T_n$ is

$$T_n = \left\lfloor \frac{N}{T} \right\rfloor \geq 4. \qquad (2)$$

The PBFT consensus process implies that the committee's number of malfunctioning nodes $f \leq \lfloor \frac{n-1}{3} \rfloor$, and EBRC is an improved algorithm based on PBFT, so EBRC also needs $f_T \leq \lfloor \frac{n-1}{3} \rfloor$ to be in the committee Reach a consensus within. Therefore, when the committee contains at least $\lfloor \frac{T_n-1}{3} \rfloor + 1$ failed nodes, the committee will fail. In worst-case scenario, the distribution of $f$ faulty nodes make each faulty committee exactly include $\lfloor \frac{T_n-1}{3} \rfloor + 1$ faulty nodes, then the blockchain network will not reach an effective consensus. Moreover, since the effectiveness of a single partition will affect the TPS (throughput) of the entire blockchain, in order to achieve the effectiveness of the consensus of the whole committee, inside the committee, the number of faulty nodes required is $f_{T_n}$:

$$f_{T_n} \leq \left\lfloor \frac{T_n - 1}{3} \right\rfloor. \qquad (3)$$

Then the number of partitions value

$$T \leq \left\lfloor \frac{N}{3 f_{T_n}} \right\rfloor. \qquad (4)$$

See (3). As T increases, the right-hand term of the inequality will fluctuate, that is, the value range of f will change, as shown in Figure 3. When $N = 400$ and $f = 10$, it can be known that the best T value should be within,[33,13] and the number of nodes in each partition is [400,33]. Secondly, considering the complexity of consensus. As we discussed earlier, the zoning structure requires consensus among multiple committees. Because the consensus time within the committee is dependent on the size of the committee, the time it takes to verify a transaction is inversely related to the number of committees involved.[34]

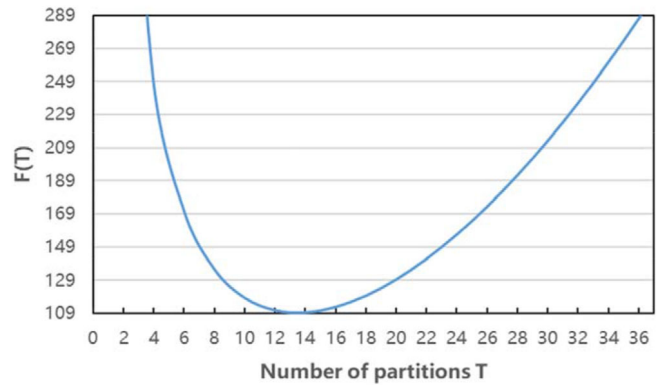

**FIGURE 4** Consensus security-performance curve ($F(T)$ performance under different $T$ values, $t_g = 0.2, t_t = 0.1$)

Therefore, another factor for selecting the best $T$ value is performance, that is, consensus complexity. The performance of committee consensus is governed by the following factors: committee internal consensus and transaction verification complexity. The committee's internal consensus uses EBRC, and its time is non-linear in the committee size, that is, $O(T_n^2)$. Transaction W is distributed among $T$ partitions in parallel, and each node in the committee works in parallel, so the transaction verification time complexity is $O\left(\frac{W}{T}\right)$. Based on the above content, the estimated theoretical running time is

$$F(T) = t_g T^2 + t_t \frac{W}{T}. \quad (5)$$

Among them, $t_g$ is the average time to achieve a consensus during the committee's internal consensus, and $t_t$ is the average time to verify the transaction on the chain. Figure 4 shows a graph of $F(T)$ and T while keeping the network size and workload unchanged. We can see that $F(T)$ decreases to a certain point before starting to increase as $(T)$ increases. At the point where the gradient is zero, the $T$-value for the best performance is provided, that is, $\frac{dF(T)}{dt}$. In this work, this curve is called the performance curve.

---

**Algorithm 1.** Optimal partition size.

---

**Input.** Node set N, partition T, transaction W.
**Output.** New Partition size.
1. **Procedure** OptimalPartitionSize(T).
2. #computes consensus security size of partition
3. $T_n = \lfloor \frac{N}{T} \rfloor, T_n \geq 4$
4. #number of faulty nodes in the committee
5. $f_{T_n} \leq \lfloor \frac{T_n - 1}{3} \rfloor >$
6. #then number of partitions is
7. $T \leq \lfloor \frac{N}{3f_{T_n}} \rfloor$
8. #compute consensus complexity
9. $F = (T) = t_g T^2 + t_t \frac{W}{T}$
10. $T = \frac{dF(T)}{dt}$
11. #combining consensus security size and complexity.
12. end procedure.

---

Finally, combine Figures 3 and 4 to obtain the optimal partition size, such as 12, by selecting the appropriate $T$ value from two different perspectives of consensus security and consensus complexity. Algorithm 1 shows the process of the optimal partition size calculation.

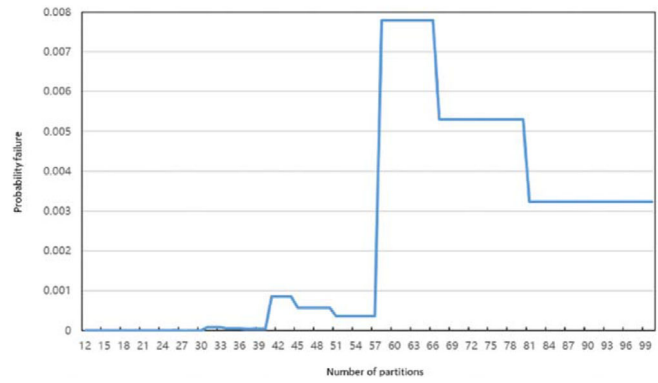

**FIGURE 5** The probability of failure P changes as the number of partitions *T* increase

In addition, according to the literature,[35] when is the total number of nodes and there are f malicious nodes, the probability that there are *k* malicious nodes in the shard using hypergeometric distribution are expressed as follows:

$$P(X=k) = \frac{\binom{f}{k}\binom{N-f}{T_n-k}}{\binom{N}{T_n}}. \tag{6}$$

Therefore, when $f = 10$ and $N = 400$, the number of partitions can also be selected according to the failure. Probability P tolerated by the system, as we can see in Figure 5.

In summary, the *T*-value for the best performance may not meet the safety standard defined in Equation (3), but Equation (3) provides multiple values that meet the safety conditions. Therefore, in combination with the failure probability, a T value that satisfies formula 3 and is closest to the optimal point on the left or right of the zero gradients on the performance curve can be selected on the performance curve.

## 4.6 | Node allocation algorithm

After the optimal partition size is obtained, the goal of this work is to divide the IoT nodes into multiple committees fairly and randomly, to realize the purpose of processing transactions in each partition independently and in parallel. In order to improve the partition's efficiency and ensure the randomness of the partition, a node allocation technique based on node reputation value is proposed in this article.

1. First, according to the calculated partition size T and the number of nodes in each partition $T_n$, when the partition size $T = 1$, no partitioning algorithm is required. Otherwise, nodes whose reputation value reaches the good level threshold are randomly assigned to each partition according to the reputation value. Among them, the nodes ranked in the top 45% of the reputation value calculate the new hash value through the unique ID and the hash value $B^{(h-1)}$ of the block agreed by the node last time, namely:

$$I(h) = H\left(B^{(h-1)}, ID\right). \tag{7}$$

I(h) will change in each round and has nothing to do with the transaction itself. When $B^{(h-1)}$ is random, then $I(h)$ is also random, and the generation of $B^{(h-1)}$ is Unpredictable and non-deterministic, so malicious nodes cannot influence the generation of $I(h)$ by changing the transaction set. Meet the random and unpredictability required by the previous article.

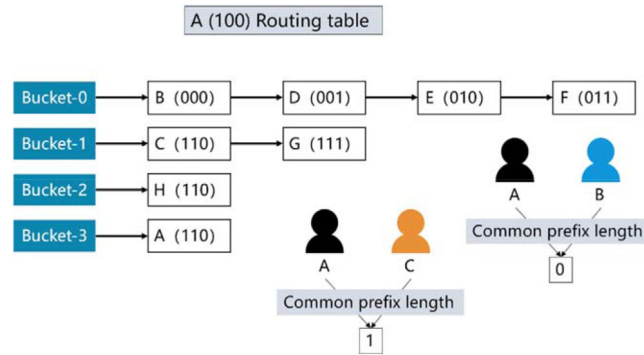

**FIGURE 6** Partition diagram

2. Then, partitions are established according to the link method of the distributed hash table,[36] as shown in Figure 6 (in the Kademlia protocol, each node obtains a bucket according to the distance between other nodes and itself, and each node in the bucket It has the same longest common prefix as the node. The larger the Bucket value, the closer the distance to the node. In the figure, the closest node to node A is node H, and the distance is 2). The node traverses the routing table, from bottom to top, Select the node closest to it to form a partition. When the number of partitions matches the number of nodes, stop the traversal to obtain the partition. Note that it is required that only nodes with higher reputation values establish partitions before nodes with lower reputation values can proceed. However, the last 55% of the nodes can directly determine the partition through the PRNG random number generator without waiting.
3. Once the partition is determined, each node broadcasts the partition number to the network so that the committee members can establish a point-to-point link.

## 4.7 | Cross-partition transaction consensus

Once the partition is completed, a committee is established. Each committee will wait for the distribution of the transaction and achieve a consensus using the EBRC consensus algorithm. The transaction's distribution is determined by the transaction's input address. Users can broadcast their transaction in the network, and the node that receives the transaction forwards it to the corresponding partition according to the input address of the transaction. In particular, it is specified here that when the transaction size is greater than $W_t$, it will be allocated to the partition with a high overall reputation value to improve transaction processing efficiency and security.

The consensus process is as shown in Section 3. The difference is that in order to enable the horizontal expansion of RSPC, we have applied the partition technology, each partition has its own partition chain, and the entire network node also has a lightweight global chain. The partition chain generates sub-blocks to include transactions generated within the partition, and the global chain generates global blocks to include the UTXO state of the entire network node during the partition reorganization. This design increases the rate of sub-block generation, thereby increasing throughput and latency.

When cross-partition transactions are generated, in order to promote cross-partition transactions, we designed a global leader node elected by all partitions to complete the cross-partition transaction consensus through a 4-stage two-way confirmation protocol, as shown in Figure 7. To facilitate cross-shard transactions, we use leader nodes to join different partitions together. This means that each partition maintains connections to other partitions in parallel, thereby enabling it to verify cross-partition transaction confirmations.

Let us suppose we have a transaction $Tx =< (A1, B1), t_s >$, $A1$ is an account in partition A, $B1$ belongs to partition B, and $t_s$ is the amount transferred from A1 to B1. It is also assumed that this transaction is sent to partition A. In order to execute this transaction, our algorithm has following steps:

1. Request Partition A sends A1 with consensus group signature to global leader G with sufficient proof of balance and transaction details $Tx =< (A1, B1), t_s >$, In a consensus period epoch, each region elects a representative node through the credit value (ranking top 50%) and VRF to form the global leader G (the number of leader nodes is required to be no less than 4). The seed $V_L$ of random number in VRF is calculated by each partition:

$$V_L = previous\_hash \bmod T_n. \tag{8}$$

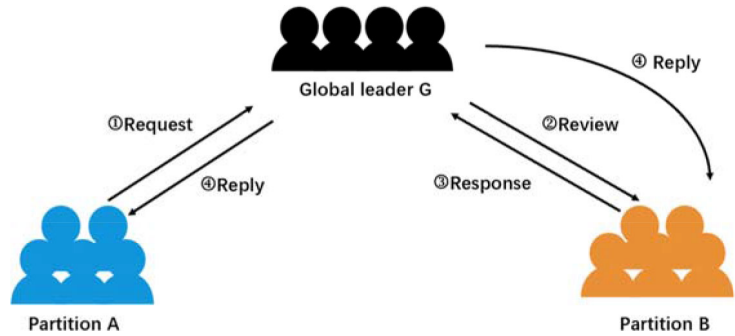

**FIGURE 7** Schematic diagram of cross-partition transactions

2. Review audit message with the global leader G signature to determine whether B partition exists.
3. Response An audit message with the global leader G signature to determine whether B partition exists.
4. Reply The global leader G sends A message confirming the correct cross-partition transaction to B and A partitions according to the results of EBRC conformance protocol, and the two partitions directly link the transactions with the signature of the global leader G without consensus. Suppose that there is a limit on the number of cross-shard transactions that can be generated between two shards. When the number of cross-shard transactions between two shards (such as partition A and partition B) reaches a certain level, it indicates that the node is not correctly partitioned. The network must be repartitioned in this case.

---

**Algorithm 2.** Cross-Partition Consensus algorithm.

---

**Input.** Node set $T_n$, partition T.
**Output.** Cross-partitioned transactions.
1. #Computer Global Leader G
2. if ($T_n > 50\%$)
3. $G = VRF(T_n)$
4. $V_l$= Previous hash mode $T_n$
5. #Cross − partitiontransactionprocess
6. $Tx =< (A1, B1), ts >$
7. $Request.Pa(< (A1B1), ts >, Psign)- > G$
8. $Audit.(< B1 > exists?, Gsign)$
9. $Response.(< B1 > exists, Gsign)$
10. $Reply.G − − > Pa(< (A1B1), ts >, Gsign)$.

---

Suppose that there is a limit on the number of cross-shard transactions that can be generated between two shards. When the number of cross-shard transactions between two shards (such as partition A and partition B) reaches a certain level, it indicates that the node is not correctly partitioned. The network must be repartitioned in this case.

In addition, the Verifiable Random Function (VRF) election stage in the EBRC consensus will determine whether to omit the operation according to the number of malicious nodes f. For example: when the number of malicious nodes f is equal to 1, the omission operation of the VRF election will be performed when the number of nodes in the committee is less. Because when the number of nodes is equal to 8, according to the election authority table, only the top 50% of the nodes in the reputation value can participate in the consensus. Under this condition, 8*50% = 4 nodes can be elected to meet the EBRC consensus conditions of. When the number of nodes is equal to 7, 7*50% ≈ 3 nodes can be elected, which does not meet the EBRC consensus conditions. Therefore, under this condition, the VRF random election link will be omitted. Through the design of the four-phase two-way confirmation protocol, we maximize the advantages of partitions, that is, allow transactions to proceed in an asynchronous and lock-free manner, to a large extent ensure the concurrency of partition transactions, and maintain the ultimate consistency of user status. Algorithm 2 shows the process of the cross-partition transaction consensus.

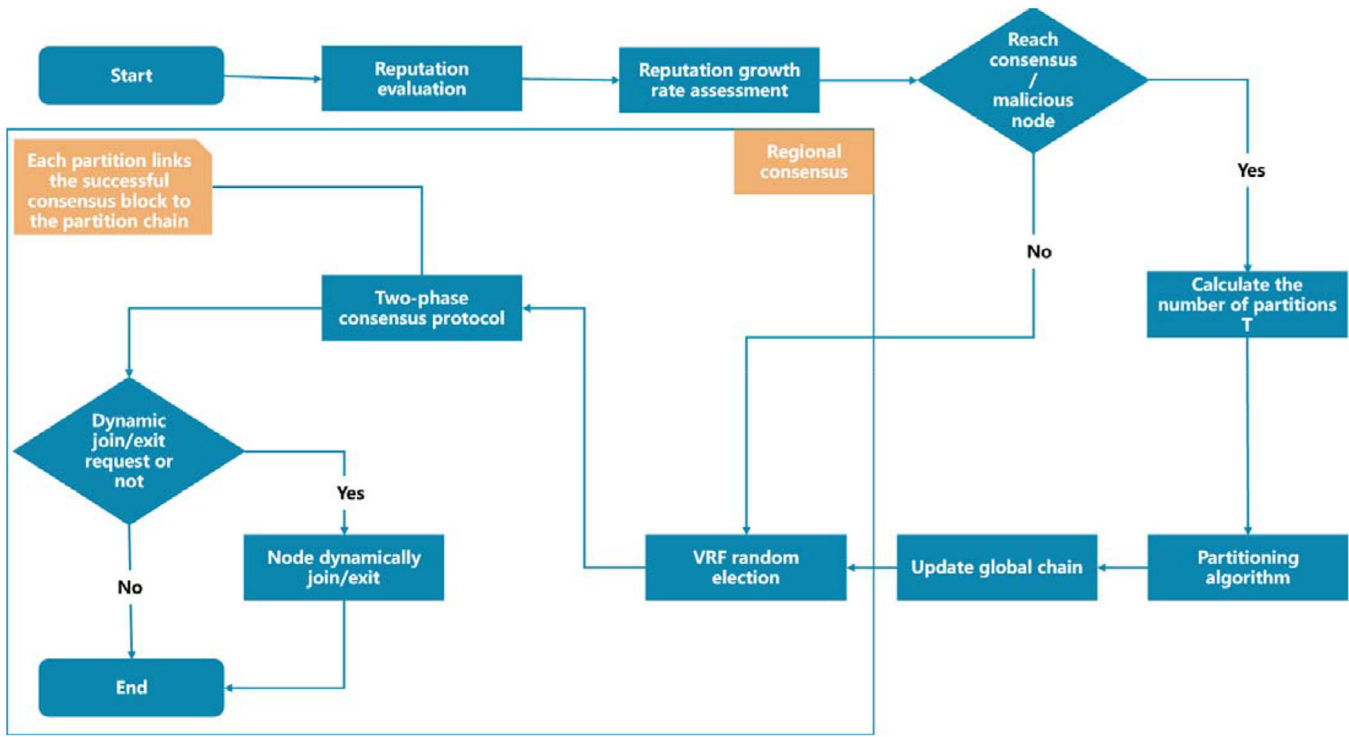

**FIGURE 8** Overall flowchart

## 4.8 | Partition reorganization strategy

The repartitioning phase allows the RSPC to reorganize its committees in response to slow-adapting opponents,[37] at the end of each epoch, it can initiate join- leave attacks[38] or destroy nodes. That is, damaged nodes can leave and rejoin the network in order to take control a committee and break the protocol's security guarantee. Moreover, in each epoch, even if no node joins/rejoins, the opponent will actively destroy a fixed number of uncorrupted nodes.

Regularly re-electing all committees is one approach to avoid this attack, faster than the opponent's ability to cause disruption. However, there is a drawback to this solution, the re-partitioning of the nodes will cause a large data migration overhead on the network. We offer a partition reorganization technique to solve this problem. That is, before repartitioning, each partition reach the consensus on the UTXO status of the current local chain and forms a state block containing the UTXO status information of each node in the partition, which is sent to the global leader G by the master node of each partition, and the global leader G agrees through EBRC A consensus is reached on the sexual agreement and the global chain is updated. After that, each node starts the transaction consensus in the new partition with the UTXO state of each node of the global chain, without the need for partition data migration.

Based on the explanations in the above sections, the overall process is sorted out as shown in Figure 8. At the same time, in practice, a ledger pruning/checkpointing mechanism, such as the mechanism described in,[19,39] can also be used to reduce storage overhead even further. For example, a significant amount of storage is usually used to keep previously used transactions.

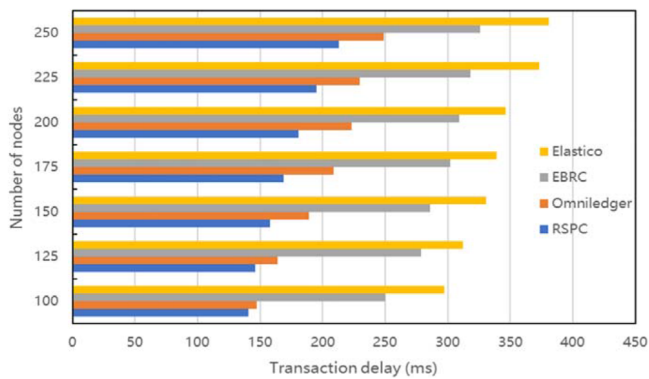

**FIGURE 9** Comparison of transaction latency between RSPC, EBRC, Elastico and Omniledger

## 5 | AGREEMENT EVALUATION

We implemented RSPC on the Hyperledger Sawtooth blockchain prototype. As the IoT network size grows, we first assess RSPC's scalability and performance. We want to make sure that RSPC's efficiency suits its theoretical analysis. We have made a comprehensive comparison with related mechanisms such as ELASTICO, OmniLedger, etc. In RSPC, we use PBFT as the consensus protocol for committees. We developed our own complete implementation of RSPC algorithm to evaluate the algorithm efficiency and effectiveness. We want to make RSPC available to the general use and keep the development running with an open-source contribution.

To test the applicability of the RPSC consensus algorithm, we devised various experiment schemes and ran them on our server machines. We use a collection of IoT nodes as 20 machines, each running multiple RSPC instances, to simulate networks of up to 1800 nodes. Each computer has an Intel Core i7 2.90 GHz processor, 16GB DRAM, and a 256GB SSD, as well as Ubuntu 16.04 and a 12-Gbps communication connection.

### 5.1 | Transaction delay

This section will examine the transaction delay of the RSPC partitioning algorithm in order to prove its effectiveness. The experiment is divided into:

1. Delay comparison of related partitioning algorithms: The experimental environment is set to divide the number of test nodes into 100, 125, 150, 175, 200, 225, and 250, and the number of malicious nodes is 5. The experimental results are shown in Figure 9.

   It can be seen from the experimental results that the transaction delay increases as the number of nodes increases in IoT crowdsensing environment but compared with EBRC, RSPC has a delay range of 102 to 213 milliseconds, and the average transaction delay is reduced by 44%. Compared with Elastico, the average transaction delay is reduced by 41%, and compared with Omniledger, the average transaction delay is reduced by 17%. This proves the effectiveness of the zoning consensus.

2. Then, this work also tested the transaction delay under different partition sizes. The experimental environment is to divide the number of test nodes into 100, 125, 150, 175, 200, 225, 250, and the number of malicious nodes to 5, and partition number is divided into 3 levels, Optimal—the calculated optimal partition size O, maximum—the most partition M1 that meets the consensus conditions, and median—the number of partitions M2 is 1/2 of the number of test nodes. Figure 10 shows the experiment results.

   From the experimental results, the transaction delay of the optimal shard size O is the smallest, and the transaction delay ranges from 141 to 199 milliseconds. This proves that the optimal shard size algorithm can ensure the maximum consensus performance and is within the safety threshold. This guarantees the validity of the consensus.

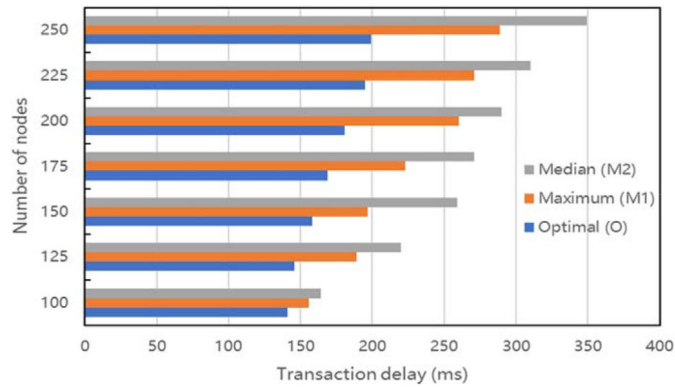

**FIGURE 10** Transaction latency under different partition sizes

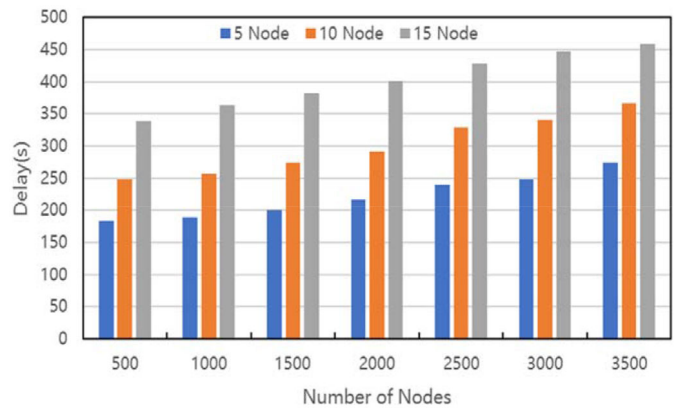

**FIGURE 11** Delay in resetting the partition

3. Partition reset delay. Figure 11 shows the delay overhead of partition reconfiguration. We measured this delay in three different scenarios, where 5, 10, or 15 nodes joined RSPC to adapt to different network sizes and committee sizes.

   As shown in Figure 11, if 15 nodes join the system instead of 5 nodes, then the reconfiguration delay will increase by approximately 1.8 times. This is because the state needs to be synchronized before the partition is reset, the added nodes still need to synchronize the global chain, and the partition size and connection time need to be calculated. In contrast, Elastico[18] cannot deal with the problem of node churn gradually and needs to reinitialize all committees. The epoch transition time of OmniLedger[19] exceeds 1000 seconds. In practice, OmniLegder's epoch transition takes more than 3 h, Because of the distributed random generating technique, it must be repeated at least 10 times to have a high chance of success. The average delay of RapidChain[20] is greater than 350, while the transition time of RSPC under the same conditions (10 nodes) is 301 s.

## 5.2 | Throughput test

In order to evaluate the impact of partitions, we measured the number of transactions executed by RSPC per second, which is an important indicator of the system's use of its main resources (ie, the number of nodes) to expand its processing capabilities. In this section we analyze the throughput performance of the algorithm in following three parts:

1. Test the throughput of the partitioning algorithm RSPC and EBRC algorithm, the number of test nodes is divided into 100, 125, 150, 175, 200, 225, 250, and the number of malicious nodes f is 20. The experimental results are shown in Figure 12:

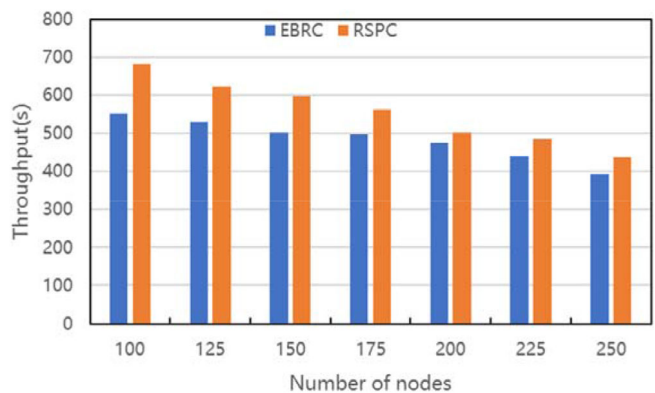

**FIGURE 12**   Comparison of transaction throughput between RSPC and EBRC algorithms

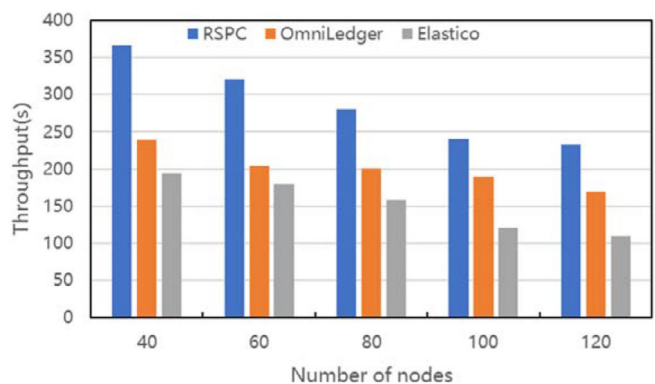

**FIGURE 13**   Comparison of transaction throughput between RSPC, ELASTICO and Omniledger

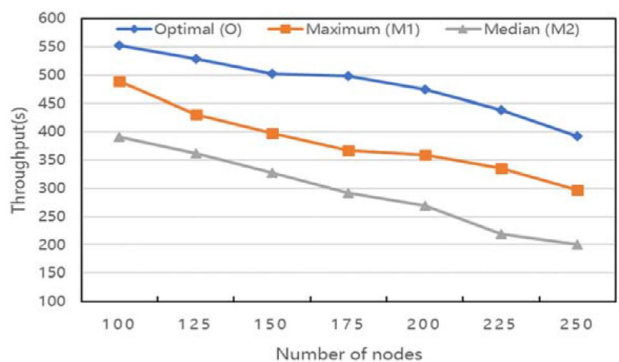

**FIGURE 14**   Transaction throughput under different partition sizes

2. To test the transaction throughput of RSPC, ELASTICO, and Omniledger with malicious nodes. The experiment is to test the transaction throughput when the no. of nodes $N = 400$ and malicious nodes f is 40, 60, 80, 100, and 120. The experiment results are shown in Figure 13.
3. To test the transaction throughput of RSPC under different shard sizes, the experimental settings are the same as the transaction delay test, and the experimental results are shown in Figure 14.

The findings of the above experiment demonstrate that with the increase in the number of malicious nodes, the transaction throughput of the system has been significantly reduced, but the transaction throughput can reach as low as 233 transactions/ms and compared with the popular partitioning algorithm, it has a significant improvement. Therefore, it is proved that the proposed RSPC partition algorithm can ensure the selection of credible committee (partition) members to ensure the stability and strength of the consensus network, thereby improving the transaction throughput of the blockchain.

**TABLE 1**  RSPC security mechanism for blockchain issues

| Security issues | Solutions |
| --- | --- |
| Integrity | SHA256 Hashing algorithm applied |
| Authorization | Nodes must be authenticated by the node public private keys through RSPC consensus in order to have privileged access to services |
| Confidentiality | Achieved using random selection of nodes for the consensus process and symmetric encryption |
| Availability | Using UTOX blocks have been placed in a right manner, so that transaction can be solved in a parallel way |
| Liveness of network | Liveness of network secured by EBRC and public private key sharing |
| User access and control | Achieved by logging transactions in a decentralized secure environment |

## 5.3 | RSPC security model analysis

In this section, we created an RSPC prototype to test its security performance for IOT environment and compared it to more advanced fragment-based protocols, it is shown in Table 1. The security architecture for an IOT crowdsensing environment must take into account the three primary security needs of Integrity, Confidentiality, and Availability. Integrity ensures that the message is sent, confidentiality ensures that only authorized users can read it, and availability ensures that the user can access each service or piece of data whenever they need it. All the nodes are guarded against malicious requests to boost IOT availability. This is accomplished by restricting the transactions that are approved to those with which each node has a common key. Nodes must first approve transactions received from the overlay before broadcasting them to all other nodes in the system. Furthermore, our RSPC framework only marginally reduces the processing times for transactions as compared to the current partitioned based IoT systems gateway solutions. For generating and distributing shared keys, there is also an additional one-time delay during initialization. In conclusion, the additional delays are negligible and have no impact on the accessibility of the IOT crowdsensing environment. We also assess the effectiveness of our solution in countering two crucial security attacks that are especially pertinent to the IOT crowdsensing environment. The first attack on IOT node is DDOS attack, in which the attacker overwhelms a specific target node by using a number of infected IoT nodes. Recent attacks that took advantage of IoT devices to execute powerful DDoS attacks have been identified. The second is a linking attack, in which the attacker creates a connection between several transactions or data ledgers using the same PK in order to discover the actual keys of linked nodes and system partitions. When choosing for a secure crowdsensing IoT implementation, the following security procedures and factors were taken into account.

## 5.4 | Data integrity and privacy

As IoT crowdsensing data travels through multiple hops in a network, a proper encryption mechanism is implemented by RSPC to ensure the confidentiality of data. Due to a diverse integration of services, devices and network, the data stored on a device is vulnerable to privacy violation by compromising nodes existing in an IoT network. The IoT devices susceptible to attacks may cause an attacker to impact the data integrity by modifying the stored data for malicious purposes which have been made secured by the RSPC implementation framework.

## 5.5 | Nodes authorization and authentication

In an IoT system, authentication between sender and receiver communicating with one another is necessary to secure data transmission between nodes. Devices must be authenticated by the node public private keys through RSPC consensus in order to have privileged access to services. The variety of different underlying architectures and ecosystems that support IoT devices are mostly to responsible for the existence of the many authentication mechanisms for IoT. These circumstances make it difficult to establish a uniform global mechanism for IoT authentication. Similar to this, authorization methods make sure that only those who have been given permission can access systems or information. A trustworthy environment that guarantees a safe communication environment was created by a suitable implementation of authorisation and authentication.

**TABLE 2** Comparison of RSPC and the most advanced sharding blockchain protocol

| Characteristic | ELASTICO | Omniledger | RapidChain | Monoxide | RSPC |
| --- | --- | --- | --- | --- | --- |
| Fault tolerance rate | 1/3 | 1/3 | 1/4 | 1/2 | 1/3 |
| Confirm the delay | 0.4 s | 0.2 s | 8 s | 15 s | 0.176 s |
| Whether to use PoW | ü | × | ü | ü | × |
| Ledger partition | × | ü | ü | ü | ü |
| Network partition | ü | ü | ü | ü | ü |
| Whether to detect malicious nodes | × | × | × | × | ü |
| Performance considerations for partitioned networks | × | × | × | × | ü |
| Partition model | Probability-Based | Probability-Based | Probability-Based | Fixed and determined | Probability-Based |
| Data migration overhead | × | ü | ü | × | × |

## 5.6 | Liveness of secured network

Through the use of traditional denial-of-service attacks, attacks on IoT devices block the delivery of services. The quality-of-service (QoS) being given to IoT nodes is deteriorated by RSPC using a range of methodologies, such as sinkhole attacks, jamming adversaries, or replay attacks that exploit IoT components at multiple tiers. High-level comparison between the results is shown in Table 2. In general, the differences between existing solutions and RSPC are as follows:

1. RSPC can evaluate nodes to generate reputation values and provides a mathematical method to calculate the size of the fragments, which can provide good system performance while ensuring that each fragment is always valid. In the stage where nodes are allocated to each partition random partitions are also performed according to the node reputation value. Therefore, the probability of all malicious nodes being assigned to the same partition is reduced, and the effectiveness of each partition is further improved.
2. In order to finalize the block. In RSPC, the effective Byzantine consensus mechanism is used to avoid the energy waste caused by the PoW consensus, and to avoid the direct use of the PBFT algorithm and the expensive 3-phase protocol (ie, pre-prepare, prepare, and commit).
3. RSPC uses a double-chain structure and proposes a novel partition reorganization strategy, which can eliminate overhead of the data migration of network nodes in the reorganization process.
4. A new four-phase two-way confirmation protocol is proposed for cross-partition transactions, allowing nodes in different partitions to process and validate cross-partition transactions.

## 6 | CONCLUSION

The IoT infrastructure relies on terminals and data aggregators having high security, low latency, and high throughput. In this article, we propose a reputation-based partition protocol (RSPC) for IoT. In comparison to existing protocols, the proposed protocol not only provides basic security features regarding IoT crowdsebsing architecture but also implements other important security characteristics. The highlight of this protocol is that it addresses the performance and reliability issues of traditional blockchain architectures. Firstly, in the case of malicious or failed nodes, the best partition size is found based on consensus security and consensus complexity problems. Secondly, in combination with the node reputation, the RSPC can guarantee that the best committees are limited by the maximum threshold of the failed node. Problem nodes can also be identified and penalized using a reputation model. The RSPC then uses a novel double-chain structure that minimizes node data migration overhead during the recombination process while also enhancing the blockchain network's horizontal scalability, resulting in a significant increase in the blockchain's throughput and performance. In addition, this work proposes a 4-stage bi-directional validation protocol to facilitate cross-shard transactions with a common data model. Finally, we implemented RSPC prototype, and the experimental results showed that the RSPC scheme was more efficient than other protocols and was able to advance the overall throughput of the blockchain while maintaining the validity of the partitioning consensus, showing better performance than before.

**ACKNOWLEDGMENTS**
The authors would like to thank Ms. Fang Luo for her help with experiments. This research is funded by China's National Key Research and Development Program 496 (Grant 2018YFC0831305).

**CONFLICT OF INTEREST**
There are no conflicts of interest declared by the authors.

**DATA AVAILABILITY STATEMENT**
The data that support the findings of this study are openly available at https://github.com/M-Zeeshan-Haider/RSPC-blockchain-partitioning-algorithm/tree/master.

**ORCID**
*Zhikui Chen*  https://orcid.org/0000-0002-9209-2189
*Shuo Yu*  https://orcid.org/0000-0003-1124-9509

**REFERENCES**
1. Qin Y, Sheng QZ, Falkner NJ, Dustdar S, Wang H, Vasilakos AV. When things matter: a survey on data-centric internet of things. *J Netw Comput Appl*. 2016;64:137-153. doi:10.1016/j.jnca.2015.12.016
2. Leijiao L, Yuqian Y, Jiaxing J, others. The content, frameworks and key Technologies of Power Internet of things for urban energy internet. *Electr Power Constr* 2019; 40(9): 91–98.
3. Banerjee M, Lee J, Choo KKR. A blockchain future for internet of things security: a position paper. *Digital Commun Netw*. 2017;4(3):149-160. doi:10.1016/j.dcan.2017.10.006
4. Reyna A, Martín C, Chen J, Soler E, Díaz M. On blockchain and its integration with IoT. Challenges and opportunities. *Fut Gener Comput Syst*. 2018;88:173-190. doi:10.1016/j.future.2018.05.046
5. Fan K, Wang S, Ren Y, et al. Blockchain-based secure time protection scheme in IoT. *IEEE Internet Things J*. 2019;6(3):4671-4679. doi:10.1109/JIOT.2018.2874222
6. Ali MS, Vecchio M, Pincheira M, Dolui K, Antonelli F, Rehmani MH. Applications of blockchains in the internet of things: a comprehensive survey. *IEEE Commun Surv Tut*. 2019;21(2):1676-1717. doi:10.1109/COMST.2018.2886932
7. Liang Y, Li Y, Shin BS. Private decentralized crowdsensing with asynchronous blockchain access. *Comput Netw*. 2022;213:109088. doi:10.1016/j.comnet.2022.109088
8. Sohrabi N, Tari Z. On the scalability of blockchain systems. *IEEE. 2020 IEEE International Conference on Cloud Engineering (IC2E)*. Sydney, Australia: IEEE; 2020:124-133.
9. Minoli D, Occhiogrosso B. Blockchain mechanisms for IoT security. *Internet Things*. 2018;1:1-13. doi:10.1016/j.iot.2018.05.002
10. Salman T, Zolanvari M, Erbad A, Jain R, Samaka M. Security services using blockchains: a state of the art survey. *IEEE Commun Surv Tut*. 2019;21(1):858-880. doi:10.1109/COMST.2018.2863956
11. Sallam A, Al Qahtani F, Gaid AS. Blockchain in internet of things: a systematic literature review. *IEEE 2021 International Conference of Technology*. Melbourne, Australia: IEEE; 2021:1-6.
12. Balasubramanium S, Sivasankar K, Rajasekaran MP. A survey on data privacy and preservation using blockchain in healthcare organization. *IEEE. 2021 International Conference on Advance Computing and Innovative Technologies in Engineering (ICACITE)*. Greater Noida, India: IEEE; 2021:956-962.
13. Secinaro S, Calandra D, Biancone P. Blockchain, trust, and trust accounting: can blockchain technology substitute trust created by intermediaries in trust accounting? A theoretical examination. *Int J Manag Pract*. 2021;14(2):129-145.
14. Vasin P. Blackcoin's proof-of-stake protocol v2. 2014; 71. https://blackcoin.co/blackcoin-pos-protocol-v2-whitepaper.pdf

15. Castro M, Liskov B. Practical byzantine fault tolerance and proactive recovery. *ACM Trans Comput Syst*. 2002;20(4):398-461. doi:10.1145/571637.571640
16. Na D, Park S. Fusion chain: a decentralized lightweight blockchain for IoT security and privacy. *Electronics*. 2021;10(4):391.
17. Jiang Q, Wan W, Qin Z, et al. Blockchain-based efficient incentive mechanism in crowdsensing. *Springer International Conference on Artificial Intelligence and Security*. Vol 13340. Qinghai, China: Springer; 2022:120-132.
18. Luu L, Narayanan V, Zheng C, Baweja K, Gilbert S, Saxena P. A secure Sharding protocol for open blockchains. *ACM. Proceedings of the 2016 ACM SIGSAC Conference on Computer and Communications Security*; 2016:17-30.
19. Kokoris-Kogias E, Jovanovic P, Gasser L, Gailly N, Syta E, Ford B. OmniLedger: a secure, scale-out, decentralized ledger via sharding. *IEEE. 2018 IEEE Symposium on Security and Privacy (SP)*. San Francisco, CA: IEEE Computer Society; 2018:583-598.
20. Zamani M, Movahedi M, Raykova M. RapidChain: scaling blockchain via full Sharding. *ACM. Proceedings of the 2018 ACM SIGSAC Conference on Computer and Communications Security*. Toronto, ON: ACM; 2018:931-948.
21. Wang J, Wang H. Monoxide: scale out blockchains with asynchronous consensus zones. *USENIX Association. 16th {USENIX} Symposium on Networked Systems Design and Implementation ({NSDI} 19)*. Boston, MA: USENIX Association; 2019:95-112.
22. Chen Z, Fiandrino C, Kantarci B. On blockchain integration into mobile crowdsensing via smart embedded devices: a comprehensive survey. *J Syst Architect*. 2021;115:102011. doi:10.1016/j.sysarc.2021.102011
23. Hu Q, Wang Z, Xu M, Cheng X. Blockchain and federated edge learning for privacy-preserving Mobile crowdsensing. *IEEE Internet Things J*. 2021. doi:10.1109/JIOT.2021.3128155
24. Perez AJ, Zeadally S. Secure and privacy-preserving crowdsensing using smart contracts: issues and solutions. *Comput Sci Rev*. 2022;43:100450. doi:10.1016/j.cosrev.2021.100450
25. Syta E, Jovanovic P, Kogias EK, et al. Scalable bias-resistant distributed randomness. *IEEE. 2017 IEEE Symposium on Security and Privacy (SP)*. San Jose, CA: IEEE Computer Society; 2017:444-460.
26. Micali S, Rabin M, Vadhan S. Verifiable random functions. *IEEE. 40th Annual Symposium on Foundations of Computer Science (Cat. No. 99CB37039)*. New York, NY: IEEE Computer Society; 1999:120-130.
27. Ning Z, Sun S, Wang X, et al. Blockchain-enabled intelligent transportation systems: a distributed crowdsensing framework. *IEEE Trans Mob Comput*. 2021;21:4201-4217.
28. Chen H, Wang Y. Sschain: a full sharding protocol for public blockchain without data migration overhead. *Pervasive and Mob Comput*. 2019;59:101055. doi:10.1016/j.pmcj.2019.101055
29. Zhang L, Li Q. Research on consensus efficiency based on practical byzantine fault tolerance. *IEEE. 2018 10th International Conference on Modelling, Identification and Control (ICMIC)*. Guiyang Xiang, China: IEEE; 2018:1-6.
30. Feng L, Zhang H, Chen Y, Lou L. Scalable dynamic multi-agent practical byzantine fault-tolerant consensus in permissioned blockchain. *Appl Sci*. 2018;8(10):1919.
31. Yuan X, Luo F, Haider MZ, Chen Z, Li Y. Efficient byzantine consensus mechanism based on reputation in IoT blockchain. *Wirel Commun Mob Comput*. 2021;2021(9952218):1-9952218, 14. doi:10.1155/2021/9952218
32. Kantesariya S, Goswami D. Determining optimal shard size in a hierarchical blockchain architecture. *IEEE. 2020 IEEE International Conference on Blockchain and Cryptocurrency (ICBC)*. Toronto, ON: IEEE; 2020:1-3.
33. Lade P, Ghosh R, Srinivasan S. Manufacturing analytics and industrial internet of things. *IEEE Intell Syst*. 2017;32(3):74-79.
34. Kumar P, Kumar R, Gupta GP, Tripathi R. A distributed framework for detecting DDoS attacks in smart contract-based blockchain-IoT systems by leveraging fog computing. *Trans Emerg Telecommun Technol*. 2021;32(6):1-31. doi:10.1002/ett.4112
35. Du M, Wang K, Liu Y, et al. Spacechain: a three-dimensional blockchain architecture for IoT security. *IEEE Wirel Commun*. 2020;27(3):38-45. doi:10.1109/MWC.001.1900466
36. Maymounkov P, Mazieres D. Kademlia: a peer-to-peer information system based on the XOR metric. *International Workshop on Peer-to-Peer Systems (IPTPS)*. Cambridge, MA: Springer; 2002:53-65.
37. Pass R, Shi E. *Hybrid consensus: efficient consensus in the Permissionless model. 91*. Vienna, Austria: Schloss Dagstuhl-Leibniz- Zentrum Fuer Informatik. 31st International Symposium on Distributed Computing (DISC 2017); 2017:39:1-39:16.
38. Douceur JR. The Sybil attack. *International Workshop on Peer-to-Peer Systems (IPTPS)*. Cambridge, MA, USA: Springer; 2002:251-260.
39. Leung D, Suhl A, Gilad Y, Zeldovich N. Vault: fast bootstrapping for cryptocurrencies. IACR Cryptol. ePrint Arch. 2018: 269.